\documentclass[12pt]{article}

\usepackage{latexsym}
\usepackage{amsfonts}
\usepackage{amssymb}
\usepackage{amsmath}
\usepackage{youngtab}

\def\R{\mathbb R}

\def\be{\begin{equation}}
\def\ee{\end{equation}}
\def\bea{\begin{eqnarray}}
\def\eea{\end{eqnarray}}
\def\beal{\begin{align*}}
\def\eeal{\end{align*}}
\def\onebox{{\text{\tiny \yng(1)}}}
\def\boxbox{{\text{\tiny \yng(2)}}}
\def\boxabox{{\text{\tiny \yng(1,1)}}}
\def\fourbox{{\text{\tiny \yng(2,2)}}}
\newcommand{\ket}[1]{\left|#1\right\rangle} 
\begin{document}

\title{Validity of the Gell-Mann formula for $sl(n, \R)$ and $su(n)$
  algebras
\hspace{.25mm}\thanks{\,This work was supported in part by MNTR,
Belgrade, Project-141036.}}
\author{\bf{Igor Salom}
\hspace{.25mm}\thanks{\,e-mail address: isalom@phy.bg.ac.rs} \\
\normalsize{Institute of Physics, P.O. Box 57, 11001 Belgrade, Serbia}
\vspace{2mm} \\
\bf{Djordje \v Sija\v cki}
\hspace{.25mm}\thanks{\,e-mail address: sijacki@phy.bg.ac.rs} \\
\normalsize{Institute of Physics, P.O. Box 57, 11001 Belgrade, Serbia}
\vspace{2mm} }

\maketitle

\begin{abstract}
The so called Gell-Mann formula, a prescription designed to
provide an inverse to the In\"on\"u-Wigner Lie algebra
contraction, has a great versatility and potential value. This
formula has no general validity as an operator expression. The
question of applicability of Gell-Mann's formula to various
algebras and their representations was only partially treated. The
validity constraints of the Gell-Mann formula for the case of
$sl(n,\R)$ and $su(n)$ algebras are clarified, and the complete
list of representations spaces for which this formula applies is
given. Explicit expressions of the $sl(n,\R)$ generators matrix
elements are obtained, in these cases, by making use of the
Gell-Mann formula.

\end{abstract}

\section{Introduction}

The Gell-Mann formula \cite{EncyclMath, HermannBook, Hermann,
Berendt} is a prescription aimed to serve as an "inverse" to the
In\"on\"u-Wigner contraction \cite{InonuWigner}. Let a symmetric Lie algebra
$\cal A = M + T$:%
\be \cal [M, M]\subset M,\quad [M, T] \subset T,\quad [T, T] \subset M,
                                                   \label{starting_algebra}
\ee %
and its In\"on\"u-Wigner contraction $\cal A' = M + U$:%
\be {\cal [M, M]\subset M,\quad [M, U] \subset U,\quad [U, U]} = \{0\},
\ee %
be given. The Gell-Mann formula prescribes that elements $T \in {\cal T}$
can be, loosely speaking, constructed as the following simple function
of the contracted algebra operators $U \in {\cal U}$ and $M \in {\cal M}$:%
\be T = i \frac{\alpha}{\sqrt{U\cdot U}}[C_2({\cal M}), U] +
\sigma U.
                                                   \label{GM_general}
\ee %
Here, $C_2({\cal M})$ denotes the second order Casimir operator of the $\cal
M$ subalgebra, $\alpha$ is a normalization constant and $\sigma$ is an
arbitrary parameter. This formula was introduced by Dothan and Ne'eman
\cite{Dothan-Neeman} and advocated by Hermann. For a mathematically strict
definition, cf. \cite{EncyclMath}.

This formula is of a great potential value due to its simplicity
and the fact that many aspects of the representation theory are
much simpler for the contracted groups/algebras (e.g.\ construction of
representations \cite{Mackey}, decompositions of a direct product of
representations \cite{HermannBook}, etc.). However, this
formula is valid, on the algebraic level, only in the case of
contraction from ${\cal A} = so(m+1,n)$ and/or ${\cal A} =
so(m,n+1)$ to ${\cal A'} = iso(m,n)$, with ${\cal M} = so(m,n)$
\cite{Sankara, Weimar}. Moreover, apart from this, the formula is
also partially applicable in a broad class of other contractions
provided one restricts to some classes of the algebra
representations. The validity of Gell-Mann's formula in a weak
sense, when an algebra representation requirement is imposed as
well, was investigated long ago by Hermann \cite{HermannBook,
Hermann}. A partial set of classes of the algebra representations
for which the Gell-Mann formula holds is listed \cite{Hermann}. No
attempt to make this list exhaustive is made, deliberately
concentrating "on what seems to be the simplest situation". This
analysis excluded, from the very beginning, the cases of
representations where the little group (in Wigner's terminology)
is non-trivially represented, not claiming a complete answer even
than.

The aim of this paper is to clarify the matters of the Gell-Mann
formula applicability for the class of $sl(n,\R)$ algebras
contracted w.r.t. their $so(n)$ maximal compact subalgebras. Note,
that owing to a direct connection of the $sl(n,\R)$ and $su(n)$
algebras, the conclusions readily convey to the latter case. Apart
from pure group-theoretical reasons, this problem is strongly
motivated by physical applications of the Gell-Mann formula to the
areas of gravity and $pD$-brane physics. The representations of
the $SL(n,\R)$ groups and their algebras, and, in particular, of
their double coverings $\overline{SL}(n,\R)$ (whose spinorial
representations are necessarily infinite-dimensional), are of
interest \cite{SijackiAffine, SijackiBranes}. In these
applications, the solution of the labeling problem of the
$\overline{SL}(n,\R)$ groups representations is only a starting
point. By a rule, a detailed information about the matrix elements
of the noncompact operators in a basis of the maximal compact
subgroup $Spin(n)$, a representation content of the $Spin(n)$
sub-representations etc. is required. The Gell-Mann formula offers
a powerful method to describe various representation details in a
simple closed analytic form.

\section{Framework}

In this paper, rather then following the approach of Hermann \cite{Hermann},
we work in the representation space of square integrable functions over the
maximal compact subgroup $Spin(n)$, with a standard invariant Haar measure:
${\cal L}^2(Spin(n))$. This representation space is large enough to provide
for all inequivalent irreducible representations of the contracted group, and
is also rich enough to contain representatives from all equivalence classes of
the $\overline{SL}(n,\R)$ group, i.e. $sl(n,\R)$ algebra, representations
\cite{HarishChandra}.

The  generators of the contracted group are generically
represented, in this space, as follows. The $so(n)$ subalgebra
operators act, in a standard way, via a right group action: %
$$
M_{ab} \ket{\phi} = -i \frac{d}{dt} \exp(i t M_{ab})\Big|_{t=0}
\ket{\phi}, \quad g' \ket{g} = \ket{g'g}, \quad \ket{\phi} \in
{\cal L}^2(Spin(n)).
$$
The Abelian operators $U_\mu$ act multiplicatively as Wigner
$D$-functions (the $SO(n)$ group matrix elements expressed as
functions of the group parameters): %
\be U_\mu \rightarrow |u| D^{\boxbox}_{x\mu}\!(g^{-1}) \equiv
\left<
\begin{array}{c} \boxbox \\ x \end{array} \right| g^{-1}
\left| \begin{array}{c} \boxbox \\ \mu \end{array} \right> ,
\label{UisD}\ee %
$|u|$ being a constant norm, $g$ being an $SO(n)$ element, and
$\boxbox$ denoting the symmetric second order tensor representation of
$SO(n)$. The $\left| \begin{array}{c} \boxbox \\ \mu \end{array} \right>$
vector from representation $\boxbox$ space is denoted by the index of the
operator $U_\mu$, whereas the vector $\left| \begin{array}{c} \boxbox \\ x
\end{array} \right>$ can be an arbitrary vector belonging to $\boxbox$ (the
choice of $x$ determines, in Wigner terminology, the little group of the
representation in question). Taking an inverse of $g$ in (\ref{UisD}) insures
the correct transformation properties.

A natural discrete orthonormal basis in the $Spin(n)$ representation space is
given by properly normalized functions of the $Spin(n)$ representation matrix
elements:%
\be \left\{\left| { \begin{array}{l@{}l}  J & \\ k & m
\end{array}} \right> \equiv \int {\scriptstyle \sqrt{dim(J)}}
D^J_{km}\!(g^{-1}) dg \ket{g}\right\}, \ \left< {
\begin{array}{l@{}l|l@{}l}  J & & J'\\ k & m & k' & m'
\end{array}} \right> = \delta_{JJ'}\delta_{kk'}\delta_{mm'},
                                          \label{naturalbasis}\ee
where $dg$ is an (normalized) invariant Haar measure. Here, $J$ stands for a
set of $Spin(n)$ irreducible representation labels, while the $k$ and $m$
labels numerate the representation basis vectors.

An action of the $so(n)$ operators in this basis is well known, and it can be
written in terms of the Clebsch-Gordan coefficients of the $Spin(n)$ group as
follows,
{\renewcommand{\arraystretch}{0.2} %
\be \left< M_{ab}\right> = \left< { \begin{array}{l@{}l} J' & \\
k' & m' \end{array}} \right| M_{ab} \left| {\begin{array}{l@{}l} J
& \\ k & m \end{array}} \right> = \delta_{JJ'} {\scriptstyle
\sqrt{C_2(J)}} \; C\!\!\!{\scriptsize
\begin{array}{c@{}c@{}c} J & \boxabox & J' \\ m & (ab) & m' \end{array}}.
                                                       \label{Maction}\ee} %
The matrix elements of the $U_\mu$ operators in this basis are
readily found to read: %
\be \left< U^{(x)}_\mu\right> = |u|\left< {
\begin{array}{l@{}l} J' & \\ k' & m' \end{array}} \right|
D^{-1\boxbox}_{x\mu} \left| {\begin{array}{l@{}l} J &
\\ k & m
\end{array}}
\right> =  |u|{\scriptstyle \sqrt{\frac{dim(J)}{dim(J')}}}
C\!\!{\scriptsize
\begin{array}{c@{}c@{}c} J & \boxbox & J' \\ k & x & k' \end{array}}
C\!\!{\scriptsize
\begin{array}{c@{}c@{}c} J & \boxbox & J' \\ m & (ab) & m' \end{array}}
\label{Uaction} . \ee %
A closed form of the matrix elements of the whole contracted algebra
$r_{\frac{n(n+1)}{2}-1}\biguplus so(n)$ (a semidirect sum of a
$\frac{n(n+1)}{2}-1$ dimensional Abelian algebra and $so(n)$)  representations
is thus explicitly given in this space by (\ref{Maction}) and (\ref{Uaction}).

Moreover, we introduce the so called, left action generators $K$ as:%
\be K_{\mu} \equiv
g^{\nu\lambda} D_{\mu\nu}^\boxabox M_\lambda, \label{Koperator}\ee %
where $g^{\nu\lambda}$ is the Cartan metric tensor of $SO(n)$. The
$K_{\mu}$ operators behave exactly as the rotation generators
$M_{\mu}$, it is only that they act on the lower left-hand side
indices of the basis (\ref{naturalbasis}):
{\renewcommand{\arraystretch}{0.2} %
\be \left< K_{ab}\right> = \left< { \begin{array}{l@{}l} J' & \\
k' & m' \end{array}} \right| K_{ab} \left| {\begin{array}{l@{}l} J
& \\ k & m \end{array}} \right> = \delta_{JJ'} {\scriptstyle
\sqrt{C_2(J)}} \; C\!\!\!{\scriptsize
\begin{array}{c@{}c@{}c} J & \boxabox & J' \\ k & (ab) & k' \end{array}}.
                                                      \label{Kaction}\ee} %
The operators $K_{\mu}$ and $M_{\mu}$ mutually commute. However, the
corresponding Casimir operators match and, in particular, we will use $\sum
K_{\mu}^2 = \sum M_{\mu}^2$ in the expression for the Gell-Mann formula
(\ref{GM_general}).

\section{Validity of the Gell-Mann formula}

The Gell-Mann formula validity problem is due to the fact that the third
commutation relation of (\ref{starting_algebra}) is not a priori satisfied as
an operator relation when the algebra elements are given by expressions
(\ref{GM_general}). In the $sl(n,\R)$ case, the $\cal T$ subspace is spanned
by $\frac 12 n(n+1) - 1$ of, so called, shear generators $T_\mu$. These
operators transform as a second order symmetric tensor w.r.t. $Spin(n)$
subgroup, and, in the Cartesian basis, satisfy: %
\be [T_{ab},T_{cd}] = i(\delta_{ac}M_{db} + \delta_{ad}M_{cb} +
\delta_{bc}M_{da} + \delta_{bd}M_{ca}).           \label{TTcommutator}\ee %

To investigate circumstances in which this relation holds, we evaluate the
commutator of two shear generators in the framework given in previous
section. The Gell-Mann formula (\ref{GM_general}) reads now: %
\be T_\mu = i\alpha [C_2(so(n))_K, D^\boxbox_{x\mu}] + \sigma
D^\boxbox_{x\mu},
                                          \label{Gell-Mann_in D terms}\ee %
where $C_2(so(n)_K) = \frac 12 \sum_{a,b=1}^n (K_{ab})^2$. By making use of
this formula, a few algebraic relations and some properties of the Wigner
$D$-functions, after some algebra we obtain:%
\bea & [T_\mu, T_\nu] = - 2\alpha^2 \big[K_{\{i},
[K_{j\}},D^\boxbox_{x\nu}]\big][K_j,D^\boxbox_{x\mu}]K_i - (\mu
\leftrightarrow \nu) & \nonumber \\%
&= \cdots= - \alpha^2 \sum_J {\sum_{\lambda, \lambda'}}
(C^{\boxbox \, \boxbox\, J}_{\ \, \mu\; \ \ \nu \ \; \lambda}
 - C^{\boxbox \, \boxbox\, J}_{\ \, \nu\; \ \ \mu \ \;
 \lambda})\cdot & \label{TT_calc}\\%
 &\Big( 2\big(C_2(J) - 2 C_2(\boxbox)\big)\big< \!\big<
 \vphantom{C}^J_{\lambda'}\big|1\otimes K_i
 \big| \vphantom{C}^\boxbox_{\ x} \big>  \big| \vphantom{C}^\boxbox_{\ x}
 \big> + &\nonumber \\%
& \big< \!\big<
 \vphantom{C}^J_{\lambda'}\big|[1\otimes K_i, C_{2(I+II)_K}]
 \big| \vphantom{C}^\boxbox_{\ x} \big>  \big| \vphantom{C}^\boxbox_{\ x}
 \big>\Big)D^J_{\lambda'\lambda}K_i,  & \nonumber
\eea%
where a summation over repeated Latin indices $i$ and $j$ that label the $K$
generators in any real basis (such that $C_{2_K} = K_iK_i$) is assumed. The
$C_{2(I+II)_K}$ operator here denotes the second order Casimir operator
acting in the tensor product of two $\boxbox$ representations,
i.e.\ $C_{2(I+II)_K} = \sum_i (K_i \otimes 1 + 1 \otimes K_i)^2$.

The summation index $J$ in (\ref{TT_calc}) runs over all
irreducible representations of the $Spin(n)$ group that appear in
the tensor product $\boxbox \otimes \boxbox$, and $\lambda,
\lambda'$ count the vectors of these representations. Since all
irreducible representations terms, apart those for which the
Clebsch-Gordan coefficient $C^{\boxbox \, \boxbox\, J}_{\ \, \mu\;
\
  \ \nu \ \; \lambda}$ is antisymmetric w.r.t. $\mu \leftrightarrow \nu$
vanish, we are left with only two values that $J$ takes: one corresponding to
the antisymmetric second order tensor $\boxabox$ and the other one
corresponding to the
representation that we denote as $\fourbox$. The fact that in the case of
$sl(n,\R)$ algebras, there is another representation term, in addition to
$\boxabox$, in the antisymmetric product of two $\boxbox$ representations
(i.e. representations that correspond to abelian $U$ operators), is in the
root of the Gell-Mann formula validity problem. Note that in the case of the
$so(m+1,n) \rightarrow iso(m,n)$, i.e. $so(m,n+1) \rightarrow iso(m,n)$)
contractions, where the Gell-Mann formula works on the algebraic level, the
contracted $U$ operators transform as $\onebox$ and the antisymmetric product
of two such representations certainly belongs to the $\boxabox$ representation
and closes upon the ${\cal M} = so(m,n)$ subalgebra.

The $so(n)$ Casimir operator values satisfy $C_2(\fourbox) = 2 C_2(\boxbox) =
4n$, implying that one of the two terms vanishes in (\ref{TT_calc}) when
$J=\fourbox$, leaving us with:%
\bea & \frac 1{2\alpha^2}[T_\mu,T_\nu] = 4(n+2) {\sum_{\lambda,
\lambda'}} C^{\boxbox \, \boxbox\, \boxabox}_{\ \, \mu\; \ \ \nu \
\; \lambda} \big< \!\big<
 \vphantom{C}^\boxabox_{\lambda'}\big|1\otimes K_i
 \big| \vphantom{C}^\boxbox_{\ x} \big>  \big| \vphantom{C}^\boxbox_{\ x}
 \big> D^\boxabox_{\lambda'\lambda}K_i - & \nonumber\\%
& {\sum_{\lambda, \lambda'}} C^{\boxbox \, \boxbox\, \boxabox}_{\
\, \mu\; \ \ \nu \ \; \lambda}
  \big< \!\big<
 \vphantom{C}^\boxabox_{\lambda'}\big|[1\otimes K_i, C_{2(I+II)_K}]
 \big| \vphantom{C}^\boxbox_{\ x} \big>  \big| \vphantom{C}^\boxbox_{\ x}
 \big>D^\boxabox_{\lambda'\lambda}K_i - & \label{TT_fin}\\%
& {\sum_{\lambda, \lambda'}} C^{\boxbox \, \boxbox\, \fourbox}_{\
\, \mu\; \ \ \nu \ \ \ \lambda}
  \big< \!\big<
 \vphantom{C}^\fourbox_{\ \lambda'}\big|[1\otimes K_i, C_{2(I+II)_K}]
 \big| \vphantom{C}^\boxbox_{\ x} \big>  \big| \vphantom{C}^\boxbox_{\ x}
 \big>D^\fourbox_{\lambda'\lambda}K_i, \nonumber \eea%
where we used that $C_2(\boxabox) = 2n-4$.

As the coefficient $\alpha$ can be adjusted freely, all that is needed
for the Gell-Mann formula to be valid is that (\ref{TT_fin}) is proportional to
the appropriate linear combination of the $Spin(n)$ generators, as determined
by the Wigner-Eckart theorem, i.e.:%
\be [T_\mu, T_\nu] \sim \sum_\lambda C^{\boxbox \, \boxbox\,
\boxabox}_{\ \mu\ \ \ \nu \ \ \lambda}M_\lambda = \sum_{\lambda,
i} C^{\boxbox \, \boxbox\, \boxabox}_{\ \mu\ \ \ \nu \ \ \lambda}
D^\boxabox_{i\lambda} K_{i}. \label{TT_CG}\ee%

An analysis of this requirement leads to a chain of conclusions
that we summarize briefly. The third term in (\ref{TT_fin}),
containing $D$ functions of the representation $\fourbox$, is to
vanish. Since it is not possible to choose vectors $x$ so that
this term vanishes identically as an operator, the remaining
possibility is to restrain the the space (\ref{naturalbasis}) of
its domain to some subspace $V = \{ \ket{v} \}$. More precisely, for this term
to  vanish, there must exist a subalgebra ${\bf L} \subset so(n)_K$,
spanned by some $\{K_\alpha\}$, such that $K_\alpha \in {\bf L}
\Rightarrow K_\alpha \ket{v} = 0$. Requiring additionally that
this subspace ought to close under an action of the shear
generators, and that the first two terms ought to yield
(\ref{TT_CG}), we arrive at the following two necessary
conditions:%

\begin{enumerate}
\item \label{req1} The algebra $\bf L$, must be a symmetric subalgebra
of $so(n)$, i.e.\
\be
\bf [L, N] \subset N, [N, N] \subset L; N=L^\perp ,
\ee
\item The vector $\left| \begin{array}{c} \boxbox \\ x
\end{array} \right>$ ought to be invariant under the $L$ subgroup action
(subgroup of $Spin(n)$ corresponding to $\bf L$), i.e.\
\be
K_\alpha \in {\bf L} \Rightarrow
K_\alpha \left| \begin{array}{c} \boxbox \\ x \end{array} \right> = 0 .
\ee%
\end{enumerate}

The space $V$ is thus $Spin(n)/L$. In Wigner's terminology, this means that
$L$ is the little group of the contracted algebra representation,
and that necessarily it is to be represented trivially. Besides, the little
group is to be a symmetric subgroup of the $Spin(n)$ group. This coincides
with one class of the solutions found by Hermann \cite{Hermann}. However, now
we demonstrated that there are no other solutions in the $sl(n,\R)$ algebra
cases, in particular, there are no solutions with little group represented
non trivially.

As for the first requirement, an inspection of the tables of
symmetric spaces, yields two possibilities: $L = Spin(m)\times
Spin(n-m)$, where $Spin(1) \equiv 1$, and, for $n=2k$, $L = U(k)$
($U$ is the unitary group). However, this second possibility
certainly does not imply another solution, since it turns out that
there is no vector satisfying the second above property.

Thus, {\it the only remaining possibility} is as follows,
\be
L = Spin(m)\times Spin(n-m) ,\quad m = 1, 2,\dots , n-1\quad Spin(1)\equiv 1.
\ee%
It is not difficult to show that proportionality of (\ref{TT_fin}) and
(\ref{TT_CG}) really holds in this case. The vector $\left| \begin{array}{c}
    \boxbox \\ x \end{array} \right>$ exists, and it is the one corresponding
to traceless diagonal $n\times n$ matrix $diag(\frac 1m, \dots, \frac 1m,
-\frac 1{n-m}, \dots, -\frac 1{n-m})$.

\section{Matrix elements}

The approach presented in this paper allows us also to write down
explicitly the matrix elements of the $sl(n,\R)$ generators in the
cases when the Gell-Mann formula is valid. The possible cases are
determined by the numbers $n$ and $m$. The corresponding
representation space (not irreducible in general) is the one over
the coset space $Spin(n)/Spin(m)\times Spin(n-m)$.
The proportionality factor $\alpha$ is determined to be: %
\be \alpha = \frac 12 \sqrt{\frac{m(n-m)}{n}}, \ee %
and, in a matrix notation for $\boxbox$ representation: %
\be \left| \begin{array}{c} \boxbox \\ x \end{array} \right> =
{\textstyle \sqrt{\frac{m(n-m)}{n}}}
diag(\frac 1m, \dots, \frac 1m,  -\frac 1{n-m}, \dots, -\frac 1{n-m}).\ee %

The Gell-Mann formula (\ref{GM_general},\ref{Gell-Mann_in D terms}), and the
matrix representation of the contracted Abelian generators $U$
(\ref{Uaction}) yield: %
\be \begin{array}{c} \left< { \begin{array}{c} J' \\
 m' \end{array}} \right| T_\mu \left| {\begin{array}{c} J
\\ m \end{array}} \right>  = \label{Taction}  \\
 {\textstyle \sqrt{\frac {m(n-m)}{4n}}\sqrt{\frac{dim(J)}{dim(J')}}}
\big(C_2(J') - C_2(J) + \sigma \big)  C\!\!{\scriptsize
\begin{array}{c@{}c@{}c} J & \boxbox & J' \\ 0 & 0 & 0 \end{array}}
C\!\!{\scriptsize
\begin{array}{c@{}c@{}c} J & \boxbox & J' \\ m & \mu & m' \end{array}}
.\end{array}\ee  %
The zeroes in the Clebsch-Gordan coefficient here denote vectors that
are invariant w.r.t.\ $Spin(m)\times Spin(n-m)$ transformations (in
that spirit $\left| \begin{array}{c} \boxbox \\ x
\end{array} \right> = \left| \begin{array}{c} \boxbox \\ 0
\end{array} \right>$). In the formula (\ref{Taction}), the space reduction
from ${\cal L}^2(Spin(n))$ to ${\cal L}^2(Spin(n)/Spin(m) \times Spin(n-m))$
implies a  reduction of the basis (\ref{naturalbasis}), i.e.
$\left|  {\begin{array}{l@{}l} J & \\ 0 & m \end{array}} \right>
\rightarrow \left| {\begin{array}{c} J \\ m \end{array}} \right>$
(only the vectors invariant w.r.t.\ left $Spin(m) \times Spin(n-m)$ action
remain).

The expression (\ref{Taction}), together with the action of the $Spin(n)$
generators (\ref{Maction}) provides an explicit form of the $SL(n,\R)$
generators representation, that is labeled by a free parameter $\sigma$. Such
representations are multiplicity free w.r.t. the maximal compact $Spin(n)$
subgroup, and all of them are tensorial.

\section{Conclusion}

In this paper, we clarified the issue of the Gell-Mann formula
validity for the $sl(n,\R)\rightarrow
r_{\frac{n(n+1)}{2}-1}\biguplus so(n)$ algebra contraction. We
have shown that the only $sl(n,\R)$ representations obtainable in
this way are given in Hilbert spaces over the symmetric spaces
$Spin(n)/Spin(m)\times Spin(n-m)$, $m = 1, 2, \dots , n-1$.
Moreover, by making use of the Gell-Mann formula in these spaces,
we have obtained a closed form expressions of the noncompact
operators (generating $SL(n,\R)/SO(n)$ cosets) irreducible
representations matrix elements. The matrix elements of both
compact and noncompact operators of the $sl(n,\R)$ algebra are
given by (\ref{Maction}) and (\ref{Taction}), respectively. In
particular, it turns out that, due to Gell-Mann's formula validity
conditions, no representations with $so(n)$ subalgebra
representations multiplicity can be obtained in this way.
Moreover, the matrix expressions  of the noncompact operators as
given by (\ref{Taction}) do not account for the $sl(n,\R)$
spinorial representations. Due to mutual connection of the
$sl(n,\R)$ and $su(n)$ algebras, the results of this paper apply
to the corresponding $su(n)$ case as well. The $SU(n)/SO(n)$
generators differ from the corresponding $sl(n,\R)$ operators by
the imaginary unit multiplicative factor, while the spinorial
representations issue in the $su(n)$ case is pointless due to the
fact that the $SU(n)$ is a simply connected (there exists no
double cover) group.

\end{document}